\documentclass{svjour3}                     
\smartqed  \usepackage{graphicx}
\usepackage{fix-cm}
\usepackage{amsmath}

\journalname{General Relativity and Gravitation}

\begin{document}

\title{Some charged polytropic models}

\author{ P. Mafa Takisa \and S. D. Maharaj }

\institute{
  P. Mafa Takisa \and S. D. Maharaj   \at
Astrophysics and Cosmology Research Unit,
School of Mathematics, Statistics and Computer Science, 
University of KwaZulu-Natal, Private Bag X54001, Durban 4000, South Africa
\\
\email{maharaj@ukzn.ac.za} \\\email{pmafatakisa@gmail.com} \\
}
\date{Received: date / Accepted: date}

\maketitle
\begin{abstract}

The Einstein-Maxwell equations with anisotropic pressures and electromagnetic field 
are studied with a polytropic equation of state. New exact solutions  to the field equations are generated in terms
of elementary functions. Special cases of the uncharged solutions of Feroze and Siddiqui (Gen Relativ Gravit 43: 1025, 2011)  and  
Maharaj and Mafa Takisa  (Gen Relativ Gravit 44: 1419, 2012) are recovered.
We also obtain exact solutions for a neutral anisotropic gravitating body for a polytrope
from our general treatment. Graphical plots indicate that the energy density, tangential pressure and anisotropy 
profiles are consistent with earlier treatments which suggest relevance
in describing relativistic compact stars.
\end{abstract}

\section{\label{b} Introduction }
In this paper we are concerned with anisotropic, charged fluids in general relativity theory satisfying the Einstein-Maxwell system. The canonical 
approach to study such a model is to specify initially the properties of matter in terms of equations of state. Then the model may be simplified by 
imposing symmetries on the spacetime manifold which eases the task of solving the field equations. The resulting family of solutions should 
be studied to confirm their physical relevance. For neutral gravitating spheres, Delgaty and Lake \cite{1a} discuss the relevant physical 
requirements and they show that only a restricted family of models satisfy the physical tests. In our approach we impose the requirement that 
the spacetime is static and spherically symmetric, specify an equation of state relating the radial pressure to the density, and choose forms for 
one of the metrics variables and the electric field. This line of approach is different from the canonical approach but has the advantage 
of simplifying the integration process. It does produce exact solutions which may be useful examples for stellar models.

The modeling of dense charged gravitating objects in strong gravitational fields has generated much 
interest in recent times because of its relevance to relativistic astrophysics. Gupta and Maurya \cite{1,2,3},
Kiess \cite{4}, Maurya and Gupta \cite{5,6,7} and Pant $et~al$ \cite{8} have generated specific charged models with desirable
physical features. These investigations require an exact solution of the Einstein-Maxwell system.
The presence of charge produces values for the redshift, luminosity and maximum mass which are different from neutral matter.
Applications of dense charged gravitating spheres include describing quarks stars, spheres with strange equation of state,
hybrid protoneutron stars, bare quark stars and the accreting process onto a compact object where the matter is likely to acquire
large amounts of electric charge as pointed out by Esculpi and Aloma \cite{9}, Sharma and Maharaj \cite{10}, Sharma and Mukherjee \cite{11,12}
and Sharma $et~al$ \cite{13} amongst others.\\

A considerable number of exact solutions to the Einstein-Maxwell system has been generated by Ivanov \cite{14}, Komathiraj and Maharaj \cite{15,16} 
and Thirukkanesh and Maharaj \cite{17} by choosing a generalized form for one of the gravitational potentials.
The solutions are represented as an infinite series in closed form in general; polynomial and algebraic functions are possible for 
particular parameter values and previously known models are regained in the appropriate limit. 
However these models do not satisfy a barotropic equation of state, relating the radial pressure to the energy density in general.
The importance of an equation of state in a stellar model has been emphasized by Varela $et~al$ \cite{18}
who provided a mechanism of dealing with anisotropic matter in a general approach.
Some solutions of the Einstein-Maxwell system found recently do in fact satisfy an equation of state.
The models of Thirukkanesh and Maharaj \cite{19}, Mafa Takisa and Maharaj \cite{20}, Thirukkanesh and Ragel \cite{21} possess a linear equation
of state for a charged anisotropic sphere. The solution of Hansraj and Maharaj \cite{22} satisfies a complicated nonlinear barotropic
equation of state with isotropic pressures. The models of Feroze and Siddiqui \cite{23} and  Maharaj and Mafa Takisa \cite{24}
satisfy a quadratic equation of state which is important in brane world models and the study of dark energy. Models with a polytropic equation
of state are rare. Thirukkanesh and Ragel \cite{21,25} have recently obtained particular uncharged models by specifying the polytropic index leading
to masses and energy densities which are consistent with observations.\\

In this paper we consider the general situation of anisotropic matter in the presence of an electromagnetic 
field satisfying a polytropic equation of state.
Our objective is to find exact solutions to the Einstein-Maxwell system. We ensure that the charge density is regular throughout the sphere
 and finite at the centre. The gravitational potential selected has a functional form which has produced physically viable
models in the past.
An advantage of our approach is that we can automatically produce a new uncharged anisotropic model, with a polytropic equation of state,
when the charge vanishes.
In Sect.~2, we express the Einstein-Maxwell system as an   equivalent set of differential equations using a transformation due to
Durgapal and Bannerji \cite{26}. In Sect.~3, we motivate the choice of the gravitational potential and the electric field intensity  
that allow us to integrate the field equations. The range of polytropic indices is considered in Sect.~4.
We obtain a family of exact solutions to the Einstein-Maxwell system for particular polytropic indices in this section. Uncharged models are also obtained.
In Sect.~5, we discuss the physical features of the model and generate graphical plots for the matter quantities. We make some closing remarks in Sect.~6.

\section{Field equations\label{Basic}}

In standard coordinates the line element for a static spherically symmetric fluid in the stellar interior has the form
\begin{equation}
\label{f1} ds^{2} = -e^{2\nu(r)} dt^{2} + e^{2\lambda(r)} dr^{2} + r^{2}(d\theta^{2} + \sin^{2}{\theta} d\phi^{2}).
\end{equation}
We are considering an anisotropic fluid in the presence of electromagnetic field; the energy momentum tensor is given by
\begin{equation}
 T_{ij}=\mbox{diag}\left[-\rho-\frac{1}{2}E^{2}, p_{r}-\frac{1}{2}E^{2}, p_{t}+\frac{1}{2}E^{2}, p_{t}+\frac{1}{2}E^{2}\right],
\end{equation}
where $\rho$ is the energy density, $p_{r}$ is the radial pressure, $p_{t}$ is the tangential pressure and $E$ is the electric field intensity.
The Einstein-Maxwell equations take the form
\begin{subequations}
\label{f2}
\begin{eqnarray}
\dfrac{1}{r^{2}}\big[r(1-e^{-2\lambda})\big]^{\prime} &=& \rho+\dfrac{1}{2}E^{2},\\
-\dfrac{1}{r^{2}}(1-e^{-2\lambda})+\dfrac{2\nu^{\prime}}{r}e^{-2\lambda} &=& p_{r}-\dfrac{1}{2}E^{2},\\
e^{-2\lambda}\bigg(\nu^{\prime\prime}+\nu^{\prime2}+\dfrac{\nu^{\prime}}{r}-\nu^{\prime}\lambda^{\prime}-\dfrac{\lambda^{\prime}}{r}\bigg) &=& p_{t}+\dfrac{1}{2}E^{2},\\
\sigma &=& \dfrac{1}{r^{2}}e^{-\lambda}(r^{2}E)^{\prime},
\end{eqnarray}
\end{subequations}
where primes represent differentiation with respect to \textit{r}, and the quantity $\sigma$ represents the proper charge density.

The fundamental equations describing the underlying gravitating model for an anisotropic charged spherically symmetric relativistic fluid are
given by the system (\ref{f2}). When the charge is absent then (\ref{f2}) is a system of three equations in five unknowns $(\nu, \lambda, \rho, 
p_{r}, p_{t})$. An uncharged solution may be generated by specifying forms for two unknowns or supplementing the system with two equations of state relating
 the matter variables as point out by Barraco $et~al$ \cite{21a}. In the presence of charge (\ref{f2}) is a system of four equations in six unknowns 
 ($\nu, \lambda, \rho, p_{r}, p_{t}, E$ or  $\sigma$). Note that if we choose a form of the electric field $E$ then the system (\ref{f2}) becomes a system of
three equations in four unknowns. A charged solution may be found by specifying forms for three unknowns or any combination of unknowns and equations of 
state relating the matter variables. The equations of state should be chosen on physical grounds. We note that the equations  (\ref{f2})  imply
\begin{equation}
\label{pedr}
 p_{r}^{\prime}=\frac{2}{r}(p_{t}-p_{r})-r(\rho+p_{r})\nu^{\prime}+\frac{E}{r^{2}}\left( r^{2}E\right)^{\prime},
\end{equation}
which is the Bianchi identity representing hydrostatic equilibrium of the charged anisotropic fluid. Equation (\ref{pedr}) indicates that the anisotropy 
 and charge influence the gradient of the pressure. These quantities may drastically affect quantities of physical importance such as surface tension 
as established by Sharma and Maharaj \cite{28a} in the generalized Tolman-Oppenheimer equation (\ref{pedr}). The specific forms of $p_{t}$ and $E$ in 
particular models studied will determine the nature of profiles of $p_{r}^{\prime}$.

We assume a polytropic equation of state relating the radial pressure $p_{r}$ to the energy density $\rho$ given by
\begin{equation}
p_{r} = \kappa\rho^{\Gamma}  \label{f7},
\end{equation}
where $\Gamma=1+(1/\eta)$ and $\eta$ is the polytropic index.

It is convenient to introduce a new 
independent  coordinate \textit{x} and introduce new metric functions \textit{y} and \textit{Z}:
\begin{equation}
\label{f6} x = Cr^2,~~ Z(x)  = e^{-2\lambda(r)}, ~~
A^{2}y^{2}(x) = e^{2\nu(r)},
\end{equation}
where \textit{A} and \textit{C} are constants. Then the equations governing the gravitational behaviour of a charged anisotropic sphere, with  
nonlinear polytropic  equation of state, are given by
\begin{subequations}
\label{f8}
\begin{eqnarray}
\label{f8a}
\dfrac{\rho}{C} &=& \frac{1-Z}{x}-2\dot{Z}-\dfrac{E^{2}}{2C},\\
\label{f8b}
p_{r} &=& \kappa\rho^{1+(1/\eta)},\\
\label{f8c}
p_{t} &=& p_{r}+\Delta,\\
\label{f8d}
\frac{\Delta}{C} &=& 4xZ \frac{\ddot{y}}{y}+\dot{Z}\left[1+ 2x\frac{\dot{y}}{y}\right]+\frac{1-Z}{x}-\dfrac{E^{2}}{C},\\
\label{f8e}
\frac{\dot{y}}{y} &=&\frac{1-Z}{4xZ}-\dfrac{E^{2}}{8CZ}+\frac{\kappa C^{1+(1/\eta)}}{4Z}\left[\frac{1-Z}{x}-2\dot{Z}-\dfrac{E^{2}}{2C}\right]^{1+(1/\eta)},\\
\label{f8f}
\dfrac{\sigma^{2}}{C} &=& \dfrac{4Z}{x}\left(x\dot{E}+E\right)^{2},
\end{eqnarray}
\end{subequations}
where $\Delta = p_t -p_r$ is called the measure of anisotropy.  The analogue of the system (\ref{f8}), with a linear equation of state, was pursued by 
Thirukkanesh and Maharaj \cite{19}. The Einstein-Maxwell equations, with a quadratic equation of state, was studied by Feroze and Siddiqui \cite{23} and 
Maharaj and Mafa Takisa \cite{24}. The system (\ref{f8}), representing gravitating matter with a polytropic equation of state, is
physically more relevant, and the model is of importance in relativistic astrophysics. However the polytropic equation of
 state is the most difficult to study because of the nonlinearity introduced through the polytropic index $\eta$. The transformed form of 
the Einstein-Maxwell equations simplifies the integration to produce exact solutions.

\section{\label{c1} Integration} 
We solve the Einstein-Maxwell field equations by choosing specific forms for the gravitational
 potential $Z$ and the electric field intensity $E$ which are physically reasonable. The model depends on obtaining a solution to (\ref{f8e}).
Equation (\ref{f8e}) becomes a first order equation in the potential $y$ which is integrable.
 
We make the choice
\begin{eqnarray}
Z &=&\frac{1+bx}{1+ax}\qquad a\neq b,\quad b\neq 0, \label{S1}
\end{eqnarray}
where $a$ and $b$ are real constants. The quantity $Z$ is regular at the stellar centre and continuous in the interior 
because of the freedom provided  by the parameters $a$ and $b$. It is important to realise that this choice for $Z$ is physically
reasonable and contains special cases of known relativistic star models. The choice (\ref{S1})
 was made by Maharaj and Mafa Takisa \cite{24} to 
generate stellar models that satisfy physical criteria for a stellar source with a quadratic equation of state.
Charged stellar models were also found by John and Maharaj \cite{27}, Thirukkanesh and Maharaj \cite{19},
Komathiraj and Maharaj \cite{28}  and Feroze and Siddiqui \cite{23} with this form of $Z$.
A detailed study of the Einstein-Maxwell system, for isotropic matter distributions, was performed by Thirukkanesh and Maharaj \cite{17}.
Neutral stellar models in general relativity have been found for special cases of the potential $Z$.
If we set $a=1$, $b=1/2$ then
we generate the Durgapal and Bannerji \cite{26} neutron star model. When $a=7$, $b=-1$ then
we generate the gravitational potential of Tikekar \cite{29} for superdense stars. Thus the form $Z$ chosen 
is likely to produce physically reasonable models for charged anisotropic spheres with a polytropic equation of state.
   
For the electric field we make the choice
\begin{eqnarray}
\frac{E^{2}}{2C}&=&\dfrac{\varepsilon x}{(1+ax)^{2}} \label{S2},
\end{eqnarray}
which has desirable physical features in the stellar interior. It is finite at the centre of the star and remains 
bounded and continuous in the interior; for large values of $x$ it approaches zero. 
A similar form of the electric field was studied by Hansraj and Maharaj \cite{22} which reduces to the uncharged Finch and Skea \cite{30}
model. Finch and Skea stars satisfy all the requirements for physical acceptability. Therefore the choice (\ref{S2}) is likely to 
produce charged anisotropic models with a polytropic equation of state.\\
  
By substituting (\ref{S1}) and (\ref{S2}) into (\ref{f8e}) we obtain the result
\begin{eqnarray}
\frac{\dot{y}}{y} &=& \frac{a-b}{4(1+bx)}-\frac{\varepsilon x}{4(1+ax)(1+bx)}\nonumber\\
& &+\frac{\kappa C^{1+(1/\eta)}(1+ax)}{4(1+bx)}\left[\frac{(a-b)(3+ax)-\varepsilon x}{(1+ax)^{2}}\right]^{1+(1/\eta)}.
  \label{S4}
\end{eqnarray}
This is a first order equation but the presence of the polytropic index $\eta$ makes it difficult to solve. The right hand side of (\ref{S4}) 
and its first derivative must be continuous to ensure integrability; clearly this is possible for a wide range of the parameters $a, b, \varepsilon$ and
$\eta$. We can integrate $(\ref{S4})$  in terms of elementary functions for particular values of $\eta$ as shown in the next section.\\

In summary the potential $Z$ and the electric field $E$ have been specified. Then the charge density must have the form
\begin{equation}
 \label{S13}
\frac{\sigma^{2}}{C} = \frac{2\varepsilon(1+bx)(3+ax)^{2}}{(1+ax)^{5}}.
\end{equation}
The energy density is given by
\begin{equation}
 \label{S8}
\dfrac{\rho}{C} =\dfrac{(a-b)(3+ax)-\varepsilon x}{(1+ax)^{2}}.
\end{equation}
On integrating (\ref{S4}) we can find the gravitational potential $y$. As $Z$ and $E$ are now known quantities, we can find the measure
of anisotropy $\Delta$ by simple substitution in (\ref{f8d}). The tangential pressure $p_{t}$ then follows from (\ref{f8c}).
Thus we must find an analytic form for $y$ to complete the integration.
\section{ Polytropic models \label{choice1}}

Newtonian polytropic models have been studied for over a hundred years.
Early results have been extensively described by Chandrasekhar \cite{31}. Particular polytropic indices have been shown to be consistent with neutron stars, 
main sequence stars, convective stellar cores of red giants and brown dwarfs, and relativistic degenerate cores of white dwarfs. When $\eta=5$ then 
the polytrope has an infinite radius, and when the index $\eta$ $\rightarrow$ $\infty$ the isothermal sphere is generated.
Polytropes have also been studied in the context of general relativity. It is important to note that in Newtonian theory polytropes with certain exponents
correspond to adiabates. The physical interpretation of the distribution in relativity is more difficult since the adiabates obey different equations 
of state as indicated in treatment of Tooper \cite{31a}. Some numerical results have  been found by Tooper \cite{32} who studied the structure of
polytropic fluid spheres for $\eta=1, 3/2, 5/2$ and $\eta=3$.
Pandey $et~al$ \cite{33} presented an exhaustive study of relativistic polytropes in the range $1/2\leq \eta\leq 3$. de Felice $et~al$ \cite{34} considered 
the structure and energy of singular general relativistic polytropes in the range $0\leq\eta\leq4.5$. Recently Thirukkanesh and Ragel \cite{21,25} found 
uncharged exact solutions with a polytropic equation of state for $\eta=1$ and $\eta=2$.
Nilsson and Uggla \cite{35}  demonstrated numerically that general relativistic perfect fluid models have finite radius for the polytropic
index $0\leq\eta\leq 3.339$. Subsequently Heinzle $et~al$ \cite{33a} performed a comprehensive dynamical systems treatment for perfect fluids
that are asymptotically polytropic. The mass-radius ratio for anisotropic matter configurations is bounded for a compact general relativistic 
object as given by Boehmer and Harko \cite{34a} and Andreasson and Boehmer \cite{35a} for general matter distributions, and they are 
consequently applicable for polytropes. In this paper we consider the polytropic index ranging over the four cases $\eta=1/2,2/3,1,2$ 
for strong gravitational fields when anisotropy and the electromagnetic field are present. The values of $\eta$ chosen produce finite models that
correspond to physically acceptable matter distributions as shown in the analyses of Pandey $et~al$ \cite{33} and  Thirukkanesh and Ragel \cite{21,25}.

\subsection{\label{c20} The case $\eta=1$}
When $\eta=1$, the equation of state $(\ref{f7})$ becomes
\begin{equation}
\label{pedro}
p_{r}=\kappa\rho^{2}.
\end{equation}
On integrating $(\ref{S4})$ we get
\begin{equation}
y=B(1+ax)^{k}[1+bx]^{l}\exp\left[F(x)\right]\label{S5},
\end{equation}
where $B$ is the constant of integration. The variable $F(x)$,
the constants $k$ and $l$ are given by
\begin{eqnarray}
F(x) &=&  \frac{C^{2}\kappa[2(2b-a)(1+ax)+(b-a)]}{2(b-a)^{2}(1+ax)^{2}}\nonumber\\
&&-\frac{C^{2}\kappa \varepsilon[4a(a-b)+\varepsilon]}{8a^{2}(a-b)(1+ax)}-\frac{C^{2}\kappa \varepsilon[2a(a^{2}-2\varepsilon)+b(2ab-\varepsilon)]}{4a^{2}(a-b)^{2}(1+ax)},\nonumber\\
k &=& C^{2}\kappa[2(a-b)]^{2}\left[\frac{b^{2}}{(b-a)^{3}}+\frac{b}{(b-a)^{2}}+ \frac{1}{4}\right]\nonumber\\
&& -\frac{2\varepsilon[(a-b)^{2}+C^{2}\kappa a\varepsilon]}{a}-4C^{2}\kappa a\varepsilon[1+b(4-3b)],\nonumber\\
l&=& \frac{(a-b)}{4b}+C^{2}\kappa[2(a-b)]^{2}\left[\frac{b^{2}}{(b-a)^{3}}+\frac{b}{(b-a)^{2}}+ \frac{1}{4}\right]\nonumber\\
&& -\frac{2\varepsilon[(a-b)^{2}+C^{2}\kappa b\varepsilon]}{b}-4C^{2}\kappa \varepsilon[(a-b)(a-3b)].\nonumber\\
\end{eqnarray}
If we set $A^{2}B^{2}=D$ and $C=1$ then the line element has the form
\begin{eqnarray}
ds^{2}&=&- D\left(1+ar^{2}\right)^{2k}(1+br^{2})^{2l}\exp[2F(r^{2})]dt^{2}+\frac{1+ar^{2}}{1+br^{2}}dr^{2}\nonumber\\
&& +r^{2}(d\theta^{2}+\sin^{2}\theta d\phi^{2})
\label{CAR1},
\end{eqnarray}
in this case.

Therefore we have obtained a new charged anisotropic model corresponding to the polytropic index $\eta$. Observe that it is possible to set $\varepsilon=0$ in this solution 
so that $E=0$ and there is no charge. Thus our approach automatically generates an uncharged model. The uncharged polytrope with $\eta=1$ is given 
by the metric
\begin{eqnarray}
ds^{2}&=& -D\left(1+ar^{2}\right)^{2\kappa[2(a-b)]^{2}\left[\frac{b^{2}}{(b-a)^{3}}+\frac{b}{(b-a)^{2}}+ \frac{1}{4}\right]}\nonumber\\
&&\times\left(1+br^{2}\right)^{2\frac{(a-b)}{4b}+\kappa [2(a-b)]^{2}\left[\frac{b^{2}}{(b-a)^{3}}+\frac{b}{(b-a)^{2}}+ \frac{1}{4}\right]}\nonumber\\
&&\times\exp\left[\frac{\kappa(2(2b-a)(1+ax)+(b-a))}{(b-a)^{2}(1+ax)^{2}}\right]dt^{2}\nonumber\\
&&+\frac{1+ar^{2}}{1+br^{2}}dr^{2}+r^{2}(d\theta^{2}+\sin^{2}\theta d\phi^{2}).\nonumber\\
\label{CARP2}
\end{eqnarray}
Note that the metric $(\ref{CAR1})$, with $\varepsilon=0$, is contained in the models of Feroze and Siddiqui \cite{23} and 
Maharaj and Mafa Takisa \cite{24}. They considered the quadratic equation of state $p_{r}=\gamma\rho^{2}+\alpha\rho+\beta$. If we set $\gamma=\kappa$, $\alpha=0$, $\beta=0$ and $E=0$ 
then we find that their solutions are equivalent to our uncharged metric $(\ref{CARP2})$.

\subsection{\label{c21} The case $\eta=2$}
When $\eta=2$, the equation of state $(\ref{f7})$ becomes
\begin{equation}
\label{pedro2}
p_{r}=\kappa\rho^{3/2}.
\end{equation}
On integrating $(\ref{S4})$ we obtain
\begin{eqnarray}
y&=&B\frac{[1+bx]^{\frac{(a-b)^{2}+\varepsilon}{4b(a-b)}}}{[1+ax]^{\frac{-\varepsilon}{4a(a-b)}}}\left[\frac{\sqrt{2a(a-b)+\varepsilon}-\sqrt{b}\sqrt{(3+ax)(a-b)-\varepsilon x}}{\sqrt{2a(a-b)+\varepsilon}+\sqrt{b}\sqrt{(3+ax)(a-b)-\varepsilon x}} \right]^{m+w}\nonumber\\
&& \times \exp[G(x)],\label{SS5}
\end{eqnarray}
where $B$ is the constant of integration. The variable $G(x)$, the constants $m$ and $w$ are given by
\begin{eqnarray}
G(x) &=& -\frac{C^{3/2}\kappa}{2(1+ax)}-\frac{C^{3}\kappa\varepsilon\sqrt{(3+ax)(a-b)-\varepsilon x}}{4a(a-b)(1+ax)},\nonumber\\
m &=& \frac{C^{3/2}\kappa[(a-b)(3b-a)+\varepsilon]^{3/2}}{2\sqrt{b}(a-b)},\nonumber\\
w &=& \frac{C^{3/2}\kappa[2a^{2}(a-b)(3a+7b)-a\varepsilon(3a+5b)]-\varepsilon^{2}(b-3a)}{4a^{3/2}(a-b)\sqrt{2a(a-b)+\varepsilon}}.\nonumber\\
\end{eqnarray}
By setting $A^{2}B^{2}=D$ and $C=1$ the line element takes the form

\begin{eqnarray}
ds^{2}&=&-D\left(1+br^{2}\right)^{\frac{(a-b)^{2}+2\varepsilon}{2b(a-b)}}\left(1+ar^{2}\right)^{\frac{\varepsilon}{2a(a-b)}}\exp[2G(r^{2})]dt^{2}\nonumber\\
&&\times \left[\frac{\sqrt{2a(a-b)+\varepsilon}-\sqrt{b}\sqrt{(3+ar^{2})(a-b)-\varepsilon r^{2}}}{\sqrt{2a(a-b)+\varepsilon}+\sqrt{b}\sqrt{(3+ar^{2})(a-b)-\varepsilon r^{2}}} \right]^{2(m+w)}\nonumber\\
&&+\frac{1+ar^{2}}{1+br^{2}}dr^{2} +r^{2}(d\theta^{2}+\sin^{2}\theta d\phi^{2})
\label{CAR2},
\end{eqnarray}
for this case.

Setting $\varepsilon=0$ implies $E=0$ and we find the uncharged polytropic model with $\eta=2$. The corresponding line element is given by 

\begin{eqnarray}
ds^{2}&=&-D\left(1+br^{2}\right)^{\frac{(a-b)^{2}}{2b(a-b)}}\exp\left[\frac{-\kappa}{(1+ar^{2})}\right]dt^{2}\nonumber\\
&&\times\left[\frac{\sqrt{2a(a-b)}-\sqrt{b}\sqrt{(3+ar^{2})(a-b)}}{\sqrt{2a(a-b)}+\sqrt{b}\sqrt{(3+ar^{2})(a-b)}} \right]^{\frac{\kappa (3b-a)\sqrt{(a-b)(3b-a)}}{\sqrt{b}}+\frac{\kappa\sqrt{a}(3a+7b)}{\sqrt{2a(a-b)}}}\nonumber\\
&&+\frac{1+ar^{2}}{1+br^{2}}dr^{2} +r^{2}(d\theta^{2}+\sin^{2}\theta d\phi^{2})
\label{CAR3},
\end{eqnarray}
which is a new solution to the Einstein-Maxwell equations with this polytropic index.

\subsection{\label{c21} The case $\eta=2/3$}
When $\eta=2/3$, the equation of state $(\ref{f7})$ is
\begin{equation}
\label{pedro3}
p_{r}=\kappa\rho^{5/2}.
\end{equation}
On integrating $(\ref{S4})$ we find
\begin{eqnarray}
y&=&B\frac{[1+bx]^{\frac{(a-b)^{2}+\varepsilon}{4b(a-b)}}}{[1+ax]^{\frac{-\varepsilon}{4a(a-b)}}}\left[\frac{\sqrt{2a(a-b)+\varepsilon}-\sqrt{b}\sqrt{(3+ax)(a-b)-\varepsilon x}}{\sqrt{2a(a-b)+\varepsilon}+\sqrt{b}\sqrt{(3+ax)(a-b)-\varepsilon x}} \right]^{p+q}\nonumber\\
&& \times \exp[H(x)],\label{SSS5}
\end{eqnarray}
where $B$ is the constant of integration. The variable $H(x)$, the constants $p$ and $q$ are given by
\begin{eqnarray}
H(x) &=& -\frac{C^{5/2}\kappa(2a(a-b)+\varepsilon)^{2}{\cal A}}{12a^{2}(a-b)(1+ax)^{3}}\nonumber\\
&& -\frac{C^{5/2}\kappa(2a(a-b)+\varepsilon)((a-b)(13a^{2}-25ab)+\varepsilon(13a+7b)){\cal A}}{48a^{2}(a-b)^{2}(1+ax)^{3}}\nonumber\\
&&-\frac{C^{5/2}\kappa[(a-b)(8a^{3}(a^{2}-\varepsilon)+4a\varepsilon(\varepsilon-1))]{\cal A}}{32a^{2}(a-b)^{3}(1+ax)}\nonumber\\
&&-\frac{C^{5/2}\kappa[-9a^{2}b^{3}+206a^{3}b^{2}(a-b^{3})+70a^{2}b^{2}(b^{2}-\varepsilon)]{\cal A}}{32a^{2}(a-b)^{3}(1+ax)}\nonumber\\
&&-\frac{C^{5/2}\kappa[a^{3}(3a^{6}-b^{3})-a\varepsilon(8a^{2}-7\varepsilon){\cal A}}{32a^{2}(a-b)^{3}(1+ax)},\nonumber\\
p &=& \frac{C^{5/2}\kappa \sqrt{b}[(a-b)(3b-a)+\varepsilon]^{5/2}}{2(a-b)},\nonumber\\
q &=& \frac{C^{5/2}\kappa[51a^{2}b^{4}\varepsilon+30a^{3}b\varepsilon^{2}+1468a^{5}b^{3}]}{32a^{5/2}(a-b)^{3}\sqrt{2a(a-b)+\varepsilon}}\nonumber\\
& & +\frac{C^{5/2}\kappa[(a+b)(498a^{4}b\varepsilon+15a^{6}s+5ab\varepsilon^{3}-5a^{8}-15a^{4}\varepsilon^{2})]}{32a^{5/2}(a-b)^{4}\sqrt{2a(a-b)+\varepsilon}}\nonumber\\
& & +\frac{C^{5/2}\kappa \varepsilon[-b(42b+75a^{4})-\varepsilon^{2}(2a+5b)+6b^{2}(a^{3}-3b^{3})]}{16a^{1/2}(a-b)^{4}\sqrt{2a(a-b)+\varepsilon}}\nonumber\\
&&  +\frac{C^{5/2} \kappa[\varepsilon^{3}(a^{3}+b^{3})+ab\varepsilon^{2}(9b^{3}+15a^{3})+535b^{4}(a^{5}+b^{5})]}{32a^{5/2}(a-b)^{4}\sqrt{2a(a-b)+\varepsilon}}\nonumber\\
&& -\frac{C^{5/2}\kappa[a^{4}b^{5}(353ab-1354)+a^{5}b(16b^{3}-85a^{3})]}{32a^{5/2}(a-b)^{4}\sqrt{2a(a-b)+\varepsilon}},\nonumber
\end{eqnarray}
where ${\cal A}=\sqrt{(3+ax)(a-b)-\varepsilon x}$. If we set $A^{2}B^{2}=D$ and $C=1$ then the line element assumes the form
\begin{eqnarray}
ds^{2}&=&-D\left(1+br^{2}\right)^{\frac{(a-b)^{2}+2\varepsilon}{2b(a-b)}}\left(1+ar^{2}\right)^{\frac{\varepsilon}{2a(a-b)}}\exp[2H(r^{2})]dt^{2}\nonumber\\
&& \times\left[\frac{\sqrt{2a(a-b)+\varepsilon}-\sqrt{b}\sqrt{(3+ar^{2})(a-b)-\varepsilon r^{2}}}{\sqrt{2a(a-b)+\varepsilon}+\sqrt{b}\sqrt{(3+ar^{2})(a-b)-\varepsilon r^{2}}} \right]^{2(p+q)}\nonumber\\
&&+\frac{1+ar^{2}}{1+br^{2}}dr^{2} +r^{2}(d\theta^{2}+\sin^{2}\theta d\phi^{2})
\label{CAR3},
\end{eqnarray}
in this case.

If we set $\varepsilon=0$ then $E=0$, and we get the uncharged polytropic model with $\eta=2/3$. The uncharged line element has the form
\begin{eqnarray}
ds^{2}&=&-D[1+br^{2}]^{\frac{(a-b)^{2}}{2b(a-b)}} \left[\frac{\sqrt{2a(a-b)}-\sqrt{b}\sqrt{(3+ar^{2})(a-b)}}{\sqrt{2a(a-b)}+\sqrt{b}\sqrt{(3+ar^{2})(a-b)}} \right]^{2(p+q)}\nonumber\\
&&\times\exp[2H(r^{2})]dt^{2}+\frac{1+ar^{2}}{1+br^{2}}dr^{2} +r^{2}(d\theta^{2}+\sin^{2}\theta d\phi^{2})
\label{CAR33},
\end{eqnarray}
which is another new model for the index  $\eta=2/3$.

\subsection{\label{c21} The case $\eta=1/2$}
When $\eta=1/2$, the equation of state $(\ref{f7})$ is
\begin{equation}
\label{pedro4}
p_{r}=\kappa\rho^{3}.
\end{equation}
On integrating $(\ref{S4})$ we find
\begin{eqnarray}
y=B(1+ax)^{s}[1+bx]^{u}\exp\left[I(x)\right],\label{SSSS5}
\end{eqnarray}
where $B$ is the constant of integration. The variable $I(x)$, the constants $s$, and $u$ are given by
\begin{eqnarray}
I(x) &=& -\frac{C^{3}\kappa(2a(a-b)+\varepsilon)^{3}}{16a^{3}(a-b)(1+ax)^{4}}-\frac{C^{3}\kappa((a-b)(a-3b)-\varepsilon)^{3}}{4(a-b)^{4}(1+ax)}\nonumber\\
&& -\frac{C^{3}\kappa(2a(a-b)+\varepsilon)^{2}[(a-b)(a(3a-5b)-2\varepsilon)-a\varepsilon]}{12a^{3}(a-b)^{2}(1+ax)^{3}}\nonumber\\ 
&& -\frac{C^{3}\kappa[6a^{4}(a^{3}+6b\varepsilon)+4a^{4}b^{2}(29a+10b)+3a\varepsilon^{3}]}{8a^{3}(a-b)^{2}(1+ax)^{2}}\nonumber\\
&& -\frac{C^{3}\kappa b\varepsilon[b\eta^{2}+3a(\varepsilon(b^{2}+3a^{2})+ab(b^{2}-a^{2}))]}{8a^{3}(a-b)^{3}(1+ax)^{2}}\nonumber\\
&& -\frac{C^{3}\kappa [36a^{4}b^{3}(a^{3}-b^{3})+12a^{2}b^{2}(\varepsilon(ab-1)-a^{2}b^{2})]}{8a^{3}(a-b)^{3}(1+ax)^{2}}\nonumber\\
&&-\frac{C^{3}\kappa a^{2}(3\varepsilon(3a-b^{2})+14b^{4})}{8a^{3}(a-b)^{3}(1+ax)^{2}},\nonumber\\
s&=& -\frac{\varepsilon[(a^{2}-b^{2})^{2}-4b(a^{2}(a-b)+b^{2})]}{4a(a-b)^{5}}\nonumber\\
&&+\frac{C^{3}\kappa[a^{2}b^{2}(a^{2}+b^{2})(136b^{2}+11a^{2})]}{4a(a-b)^{5}}\nonumber\\
&&-\frac{C^{3}\kappa [9ab^{5}(3a^{2}-19b^{2})+3b^{2}\varepsilon^{2}(4a^{2}+3b)]}{4(a-b)^{5}}\nonumber\\
&&+\frac{C^{3}\kappa[3ab^{4}\varepsilon(4a+9b)+ab\varepsilon(a^{3}b+\varepsilon^{2})]}{4(a-b)^{5}}\nonumber\\
&& -\frac{C^{3}\kappa a^{2}b[a^{2}+17ab^{2}-22b\varepsilon]}{4(a-b)^{4}}\nonumber\\
&&-\frac{C^{3}\kappa[3a^{3}b\varepsilon(a^{2}+1)+40a^{3}b^{3}(a^{2}-\varepsilon)]}{4(a-b)^{5}},\nonumber\\
u &=& \frac{(a^{2}+b^{2})(4ab\varepsilon+15a^{2}b{2})-\varepsilon(a^{4}+b^{4})}{4b(a-b)^{5}}\nonumber\\
&&+\frac{(a^{3}-b^{3})-6ab(a^{4}+b^{4}+ab(3+\varepsilon))}{4b(a-b)^{5}}\nonumber\\
&& +\frac{C^{3}\kappa[12a^{3}b^{3}(a^{2}-b^{2})+b^{2}\varepsilon(6ab-\varepsilon^{2})]}{4b(a-b)^{4}}\nonumber\\
&&+\frac{C^{3}\kappa[27b^{3}(1+b^{3})+a^{3}b^{2}(a^{3}-4b^{3})]}{4b(a-b)^{4}},\nonumber\\
&& +\frac{C^{3}\kappa[ab^{4}(57a^{2}+108b^{2})-3b^{2}\varepsilon^{2}(1+3b)-3ab^{3}(21ab^{2}+22b\varepsilon)]}{4b(a-b)^{4}},\nonumber
\end{eqnarray}
in this case.
If we set $A^{2}B^{2}=D$ and $C=1$ then the line element is given by
\begin{eqnarray}
ds^{2}&=& -D(1+ar^{2})^{2s}(1+br^{2})^{2u}\exp[2I(r^{2})]dt^{2}+\frac{1+ar^{2}}{1+br^{2}}dr^{2}\nonumber\\
&& +r^{2}(d\theta^{2}+\sin^{2}\theta d\phi^{2}),
\label{CAR11}
\end{eqnarray}
for this case.

Setting $\varepsilon=0$ implies $E=0$, and we generate the uncharged polytropic model with $\eta=1/2$. The corresponding line element is given by 
\begin{eqnarray}
ds^{2}&=& -D(1+ar^{2})^{2s}(1+br^{2})^{2u}\exp[2I(r^{2})]dt^{2}+\frac{1+ar^{2}}{1+br^{2}}dr^{2}\nonumber\\
&& +r^{2}(d\theta^{2}+\sin^{2}\theta d\phi^{2})
\label{CAR111},
\end{eqnarray}
which is a new solution to the Einstein-Maxwell equations.
\section{\label{d} Physical Analysis}
In this section we indicate that the exact polytropic solutions found in Sect. 4 are physically reasonable. The gravitational potential $Z$ is regular 
at the centre and well behaved in the interior. The potentials $y$ presented for various cases in Sect. 4 are given in terms of simple elementary functions.
They are regular at the stellar centre and continuous in the interior. The potentials $Z$ and $y$ reduce for particular values of parameters to 
relativistic stellar models studied previously which have been shown to possess desirable physical features. Clearly the choice of the  electric field $E$ in
$(\ref{S2})$ is physically acceptable as shown by Hansraj and Maharaj \cite{22}. The choice of $E$ leads to forms of charge density $\sigma$ in 
$(\ref{S13})$ and the energy density $\rho$ in $(\ref{S8})$ given in terms of rational functions. The quantities $E$, $\sigma$ and $\rho$ become 
decreasing functions for large values of $x$.

We used the programming language Python to generate two sets of plots for the radial pressure $p_{r}$, the tangential pressure $p_{t}$, and the anisotropy
$\Delta$ for the polytropic indices $\eta=1/2, 2/3, 1, 2$. These represent profiles for charged anisotropic matter with $\varepsilon\neq0$ for
$a=5.5$, $b=3.0$, $\varepsilon=1$, the boundary $r=4$, $C=1$ and $\kappa$ given by the causality condition 
$\frac{dp_{r}}{d\rho}\leq1$ for each case. In the first set of figures, we have plotted $p_{r}$, $p_{t}$ and $\Delta$ against the radial coordinate $r$: Fig. 1 represents
the radial pressure, Fig. 2 represents the tangential pressure, and Fig. 3 represents the anisotropy. The radial pressure is a finite and decreasing
function in Fig. 1. The tangential pressure in Fig. 2 initially increases, reaches a maximum and then decreases. The anisotropy in Fig. 3 also reaches
a maximum in the interior and then decreases. These profiles are similar to other studies. The high values of $p_{t}$ in central regions of a star is 
reasonable  as pointed out by Karmakar  $et~al$ \cite{38} because of conservation of angular momentum in quasi-equilibrium contraction of a compact body.
The profile of $\Delta$ is similar to the profiles generated in studies of strange stars with quark matter by Sharma and Maharaj \cite{10} and
Tikekar and Jotania \cite{39}.
In the second set of figures, we have plotted $p_{r}$, $p_{t}$ and $\Delta$ against the density $\rho$: Fig. 4 represents radial pressure, Fig. 5 
represents the tangential pressure, and Fig. 6 represents the anisotropy. We have utilized the forms for $p_{r}$ from Sect. 4 and the functions
for $p_{t}$ and $\Delta$ listed in the Appendix. The radial pressure pressure remains an increasing function in Fig. 4. The tangential pressure $p_{t}$
increases to a maximum and then becomes a decreasing function in Fig. 5. This feature is to be expected as we commented above about the expected higher
values of $p_{t}$ in the central regions. In the same way, in Fig. 6, the anisotropy reaches a maximum in the interior and then decreases. Our profiles are
similar to those given by Ray  $et~al$ \cite{36} who showed that the presence of electric charge has a significant effect on the phenomenology of compact
stars with intense gravitational fields. Observe that the profiles of the radial pressure $p_{r}$ increases as a function of the energy in Fig. 4 for 
each polytropic index. The gradient is larger as the polytropic index increases; the behaviour is consistent with the physical requirements of 
Pandey $et~al$ \cite{33}. We observe the same behaviour for the profiles for $p_{t}$ and $\Delta$. Finally in Fig. 7 we have plotted the speed of sound
$\frac{dp_{r}}{d\rho}$. This quantity is always less than unity and the causality is maintained which is a requirement  for a physical object as 
indicated by Delgaty and Lake \cite{1a}.

\begin{figure}
\vskip .2cm \centering
\includegraphics[angle = 0,scale = 0.20]{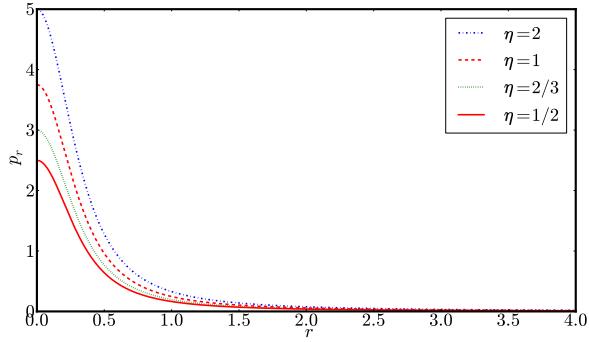}
\caption{Radial pressure $p_{r}(r)$ }
\label{fig1}
\end{figure}

\begin{figure}
\vskip .2cm \centering
\includegraphics[angle = 0,scale = 0.20]{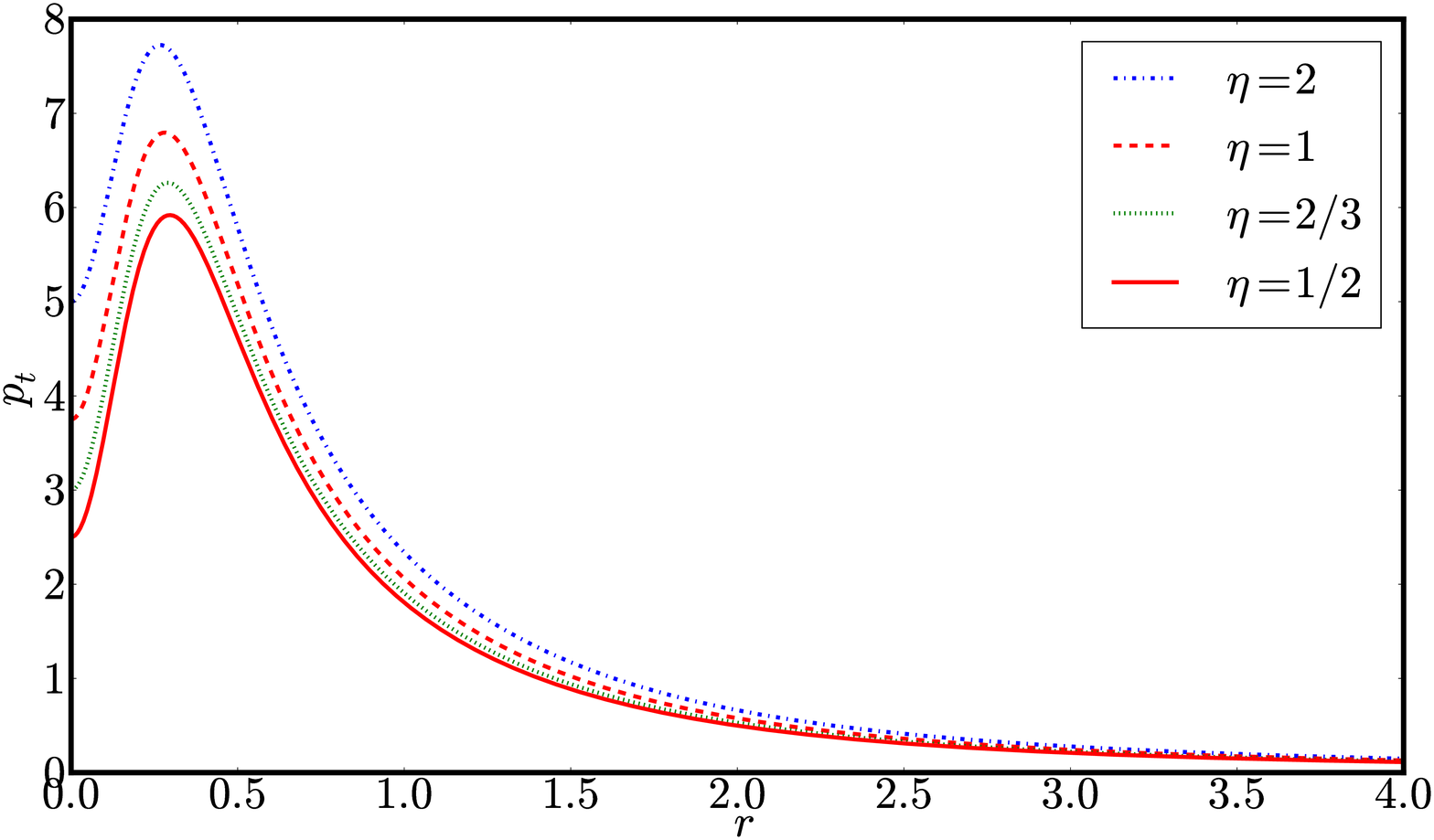}
\caption{Tangential pressure $p_{t}(r)$ }
\label{fig2}
\end{figure}

\begin{figure}
\vskip .2cm \centering
\includegraphics[angle = 0,scale = 0.20]{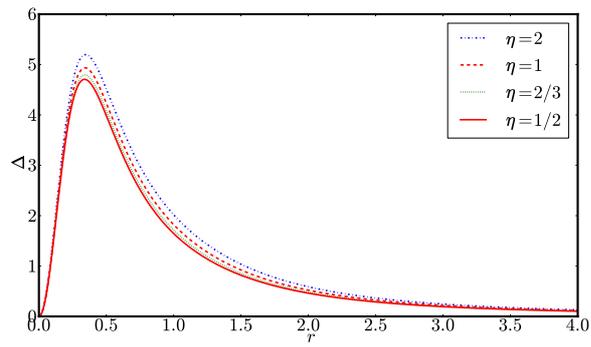}
\caption{Anisotropy $\Delta(r)$ }
\label{fig3}
\end{figure}

\begin{figure}
\vskip .2cm \centering
\includegraphics[angle = 0,scale = 0.20]{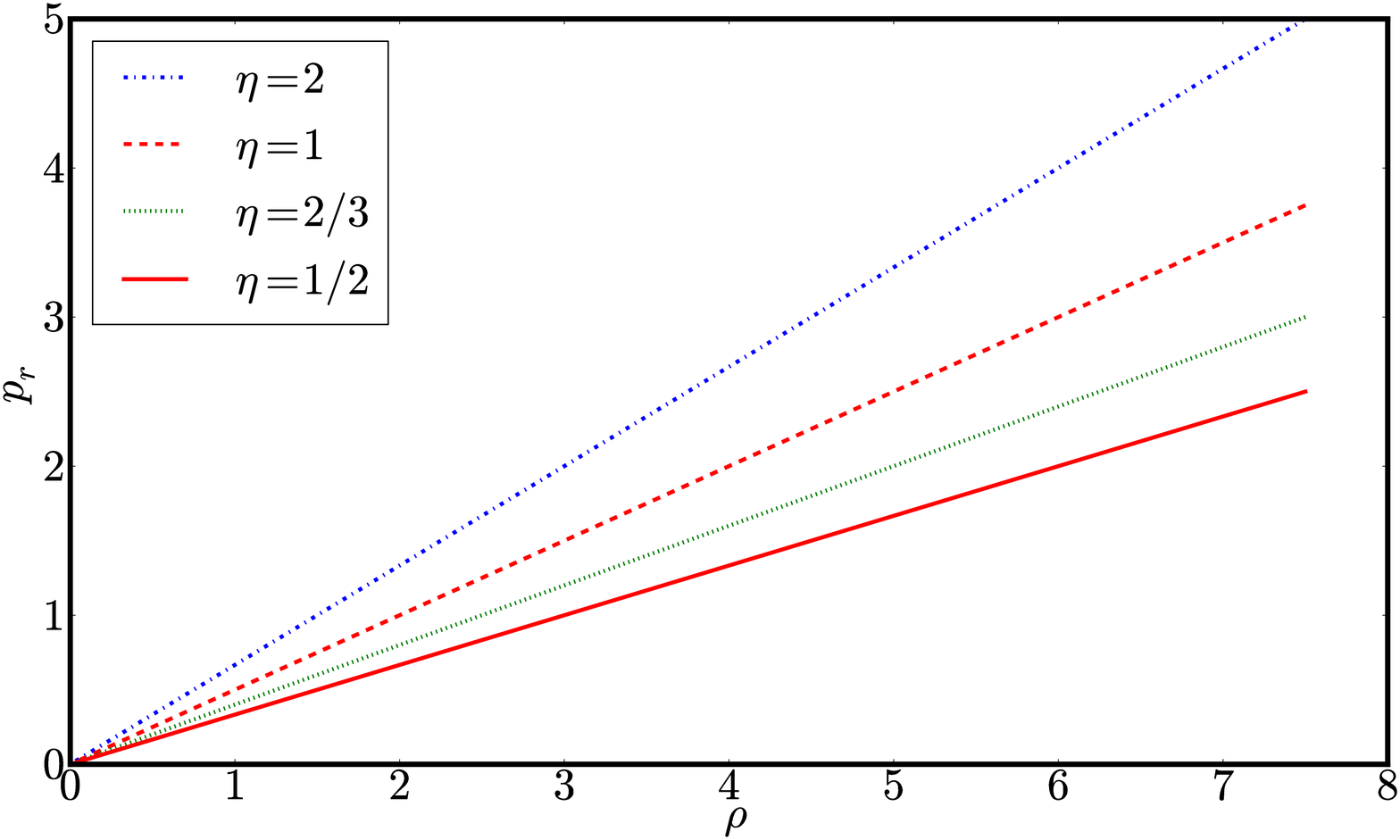}
\caption{Radial pressure $p_{r}(\rho)$ }
\label{fig4}
\end{figure}

\begin{figure}
\vskip .2cm \centering
\includegraphics[angle = 0,scale = 0.20]{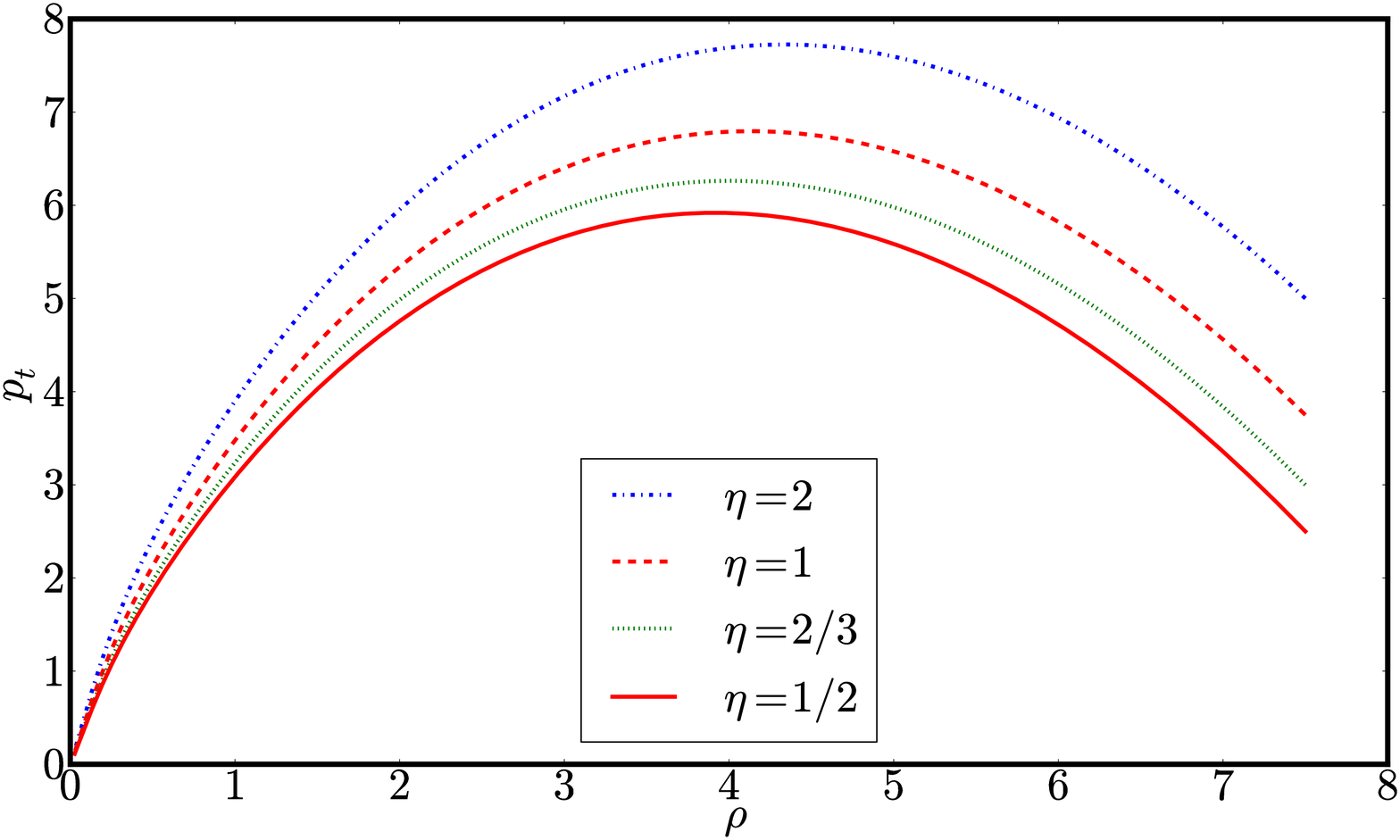}
\caption{Tangential pressure $p_{t}(\rho)$ }
\label{fig5}
\end{figure}

\begin{figure}
\vskip .2cm \centering
\includegraphics[angle = 0,scale = 0.20]{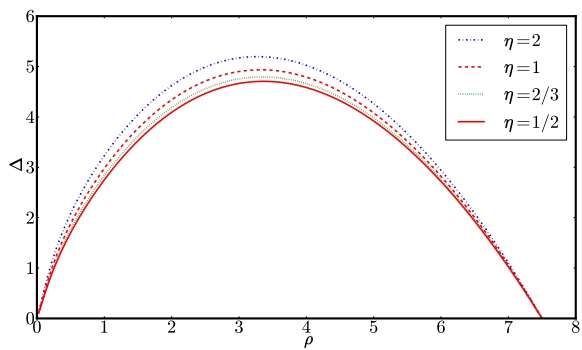}
\caption{Anisotropy $\Delta(\rho)$}
\label{fig6}
\end{figure}

\begin{figure}
\vskip .2cm \centering
\includegraphics[angle = 0,scale = 0.20]{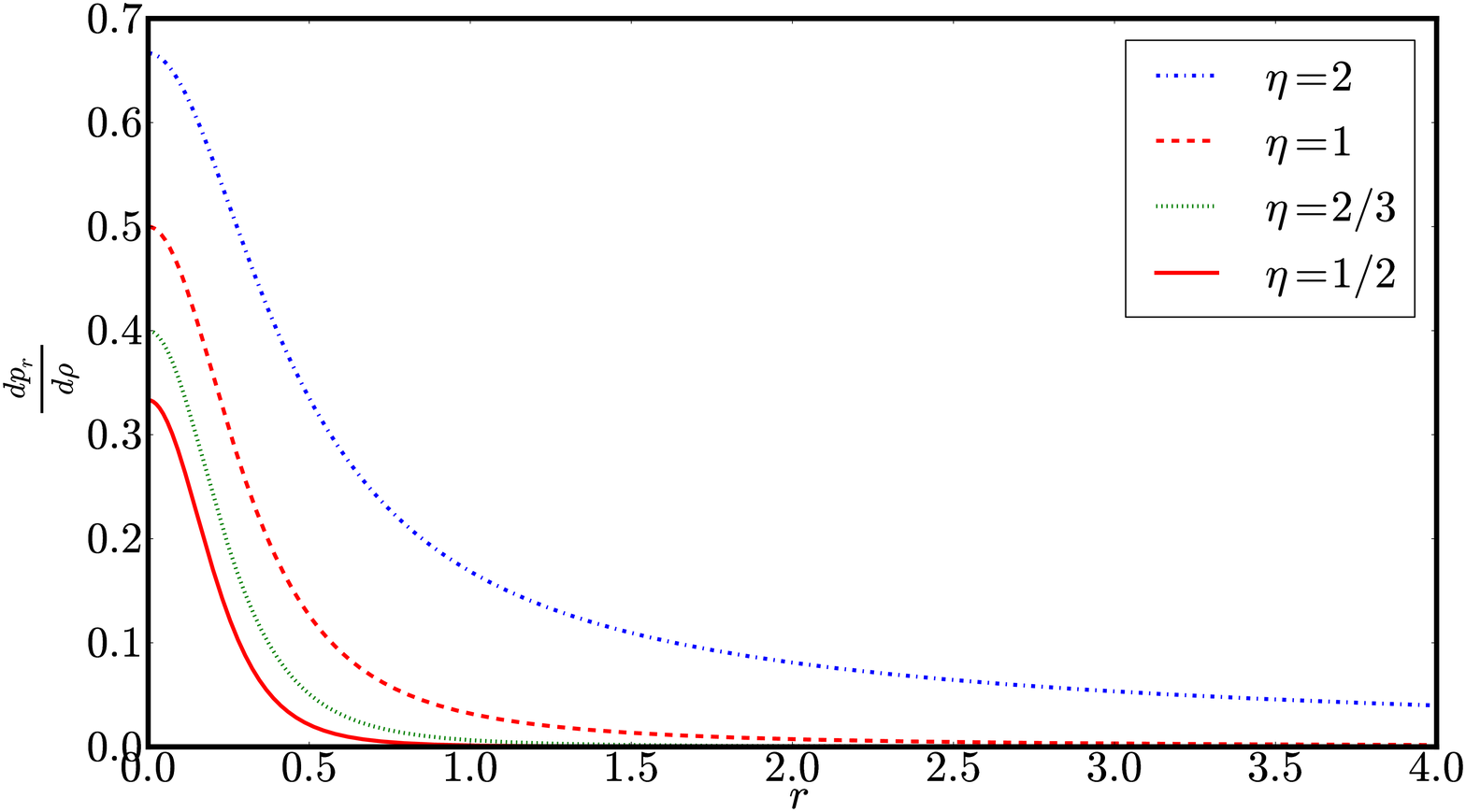}
\caption{Speed of sound $\frac{dp_{r}}{d\rho}(r)$}
\label{fig7}
\end{figure}

\section{\label{e} Discussion}
In this paper we have generated new exact solutions to the Einstein-Maxwell system of equations with a polytropic equation of state.
These solutions may be used to model compact objects which are anisotropic and charged. Note that the solutions are expressed in terms of elementary
functions which facilitate a physical study. A graphical analysis shows that the gravitational potentials and matter variables are regular at the 
centre and well behaved in the interior. It would be interesting to relate these new solutions to particular astronomical objects such as 
SAX J1804.4-3658 as was done by Dey $et~al$ \cite{40,41,42}, in the absence of charge, and Mafa Takisa and Maharaj \cite{20}, in the presence of charge.
Such a study will reinforce the astrophysical significance of the models in this paper. We point out that our approach automatically leads to new uncharged
anisotropic solutions when the electric field  $E=0$.

\newpage
\appendix
\section*{Appendix}
In the appendix we list expressions for tangential pressure $p_{t}$ and the measure of anisotropy $\Delta$ for each polytropic index considered in this paper. 
These quantities assist in the generation of graphical plots in the physical analysis. These quantities are given by

\subsection*{(a) For index  $\eta=1$:}
\begin{subequations}
\begin{eqnarray}
\label{helene}
p_{t} &=&  \frac{4xC(1+bx)}{1+ax}\left[\frac{k(k-1)a^{2}}{(1+ax)^{2}}
+\frac{2klab}{(1+ax)(1+bx)} + \frac{2ka\dot{F}(x)}{1+ax}+\frac{b^{2}l(l-1)}{(1+bx)^{2}}\right.\nonumber\\
& & \left.  + \frac{2lb\dot{F}(x)}{1+bx}+ \ddot{F}(x)+ \dot{F}(x)^{2}\right]+ 2xC\left[\frac{ak}{1+ax}+ \frac{b}{1+bx} + \dot{F}(x) \right]\nonumber\\
&&+\frac{C(a-b)ax-2\varepsilon x}{(1+ax)^{2}}+\kappa C^{2}\left[\dfrac{(a-b)(3+ax)-\varepsilon x}{(1+ax)^{2}}\right]^{2},\\
\label{helene1}
\Delta &=&  \frac{4xC(1+bx)}{1+ax}\left[\frac{k(k-1)a^{2}}{(1+ax)^{2}}
+\frac{2klab}{(1+ax)(1+bx)} + \frac{2ka\dot{F}(x)}{1+ax}+\frac{b^{2}l(l-1)}{(1+bx)^{2}}\right.\nonumber\\
& & \left.  + \frac{2lb\dot{F}(x)}{1+bx}+ \ddot{F}(x)+ \dot{F}(x)^{2}\right]+ 2xC\left[\frac{ak}{1+ax}+ \frac{b}{1+bx} + \dot{F}(x) \right]\nonumber\\
&&+\frac{C(a-b)ax-2\varepsilon x}{(1+ax)^{2}}.
\end{eqnarray}
\end{subequations}

\subsection*{(b) For index  $\eta=2$:}
\begin{subequations}
\begin{eqnarray}
\label{helene12}
p_{t} &=& \frac{4xC(1+bx)}{1+ax}\left[\frac{d}{dx}\left(\frac{b((a-b)^{2}+\varepsilon)}{4b(a-b)(1+bx)}-\frac{a\varepsilon}{4a(a-b)(1+ax)}\right. \right.\nonumber\\
 &&\left. \left. -\frac{(m+w)\sqrt{b}(a(a-b)\varepsilon)}{2T(\sqrt{2a(a-b)+\varepsilon}+\sqrt{b}T)}\right)+\frac{\dot{y}^{2}}{y^{2}}\right]+\frac{C(b-a)}{(1+ax)^{2}}\nonumber\\
&&\times\left[1+2x\left(\frac{b((a-b)^{2}+\varepsilon)}{4b(a-b)(1+bx)}-\frac{a\varepsilon}{4a(a-b)(1+ax)}\right. \right.\nonumber\\
&&\left. \left.-\frac{(m+w)\sqrt{b}(a(a-b)\varepsilon)}{2T(\sqrt{2a(a-b)+\varepsilon}+\sqrt{b}T)}\right)  \right]\nonumber\\
&&+\frac{(a-b)(1+ax)-2\varepsilon x}{(1+ax)^{2}}+\kappa C^{3/2}\left[\dfrac{(a-b)(3+ax)-\varepsilon x}{(1+ax)^{2}}\right]^{3/2},\\
\label{helene22}
\Delta &=&\frac{4xC(1+bx)}{1+ax}\left[\frac{d}{dx}\left(\frac{b((a-b)^{2}+\varepsilon)}{4b(a-b)(1+bx)}-\frac{a\varepsilon}{4a(a-b)(1+ax)}\right. \right.\nonumber\\
 &&\left. \left. -\frac{(m+w)\sqrt{b}(a(a-b)\varepsilon)}{2T(\sqrt{2a(a-b)+\varepsilon}+\sqrt{b}T)}\right)+\frac{\dot{y}^{2}}{y^{2}}\right]+\frac{C(b-a)}{(1+ax)^{2}}\nonumber\\
&&\times\left[1+2x\left(\frac{b((a-b)^{2}+\varepsilon)}{4b(a-b)(1+bx)}-\frac{a\varepsilon}{4a(a-b)(1+ax)}\right. \right.\nonumber\\
&&\left. \left.-\frac{(m+w)\sqrt{b}(a(a-b)\varepsilon)}{2T(\sqrt{2a(a-b)+\varepsilon}+\sqrt{b}T)}\right)  \right]+\frac{(a-b)(1+ax)-2\varepsilon x}{(1+ax)^{2}},
\end{eqnarray}
\end{subequations}
where $T=\sqrt{(3+ax)(a-b)-\varepsilon x}$.

\subsection*{(c) For index $\eta=2/3$:}
\begin{subequations}
\begin{eqnarray}
\label{helene122}
p_{t} &=& \frac{4xC(1+bx)}{1+ax}\left[\frac{d}{dx}\left(\frac{b((a-b)^{2}+\varepsilon)}{4b(a-b)(1+bx)}-\frac{a\varepsilon}{4a(a-b)(1+ax)}\right. \right. \nonumber\\
 &&\left. \left. -\frac{(p+q)\sqrt{b}(a(a-b)\varepsilon)}{2T(\sqrt{2a(a-b)+\varepsilon}+\sqrt{b}T)}\right)+\frac{\dot{y}^{2}}{y^{2}}\right]+\frac{C(b-a)}{(1+ax)^{2}}\nonumber\\
&&\times\left[1+2x\left(\frac{b((a-b)^{2}+\varepsilon)}{4b(a-b)(1+bx)}-\frac{a\varepsilon}{4a(a-b)(1+ax)}\right. \right. \nonumber\\
&&\left. \left.-\frac{(p+q)\sqrt{b}(a(a-b)\varepsilon)}{2T(\sqrt{2a(a-b)+\varepsilon}+\sqrt{b}T(x))}\right)  \right]\nonumber\\
&&+\frac{(a-b)(1+ax)-2\varepsilon x}{(1+ax)^{2}}+\kappa C^{5/2}\left[\dfrac{(a-b)(3+ax)-\varepsilon x}{(1+ax)^{2}}\right]^{5/2},\\
\label{helene222}
\Delta &=&\frac{4xC(1+bx)}{1+ax}\left[\frac{d}{dx}\left(\frac{b((a-b)^{2}+\varepsilon)}{4b(a-b)(1+bx)}-\frac{a\varepsilon}{4a(a-b)(1+ax)}\right. \right.\nonumber\\
 &&\left. \left. -\frac{(p+q)\sqrt{b}(a(a-b)\varepsilon)}{2T(\sqrt{2a(a-b)+\varepsilon}+\sqrt{b}T)}\right)+\frac{\dot{y}^{2}}{y^{2}}\right]+\frac{C(b-a)}{(1+ax)^{2}}\nonumber\\
&&\times\left[1+2x\left(\frac{b((a-b)^{2}+\varepsilon)}{4b(a-b)(1+bx)}-\frac{a\varepsilon}{4a(a-b)(1+ax)}\right. \right. \nonumber\\
&&\left. \left.-\frac{(p+q)\sqrt{b}(a(a-b)\varepsilon)}{2T(\sqrt{2a(a-b)+\varepsilon}+\sqrt{b}T)}\right)  \right] +\frac{(a-b)(1+ax)-2\varepsilon x}{(1+ax)^{2}}.
\end{eqnarray}
\end{subequations}

\subsection*{(d) For index  $\eta=1/2$:}
\begin{subequations}
\begin{eqnarray}
\label{helene1}
p_{t} &=&  \frac{4xC(1+bx)}{1+ax}\left[\frac{s(s-1)a^{2}}{(1+ax)^{2}}
+\frac{2suab}{(1+ax)(1+bx)} + \frac{2sa\dot{I}(x)}{1+ax}+\frac{b^{2}u(u-1)}{(1+bx)^{2}}\right.\nonumber\\
& & \left.  + \frac{2ub\dot{I}(x)}{1+bx}+ \ddot{I}(x)+ \dot{I}(x)^{2}\right]+ 2xC\left[\frac{as}{1+ax}+ \frac{b}{1+bx} + \dot{I}(x) \right]\nonumber\\
&&+\frac{C(a-b)ax-2\varepsilon x}{(1+ax)^{2}}+\kappa C^{2}\left[\dfrac{(a-b)(3+ax)-\varepsilon x}{(1+ax)^{2}}\right]^{3},\\
\label{helene1}
\Delta &=&  \frac{4xC(1+bx)}{1+ax}\left[\frac{s(s-1)a^{2}}{(1+ax)^{2}}
+\frac{2suab}{(1+ax)(1+bx)} + \frac{2sa\dot{I}(x)}{1+ax}+\frac{b^{2}l(u-1)}{(1+bx)^{2}}\right.\nonumber\\
& & \left.  + \frac{2lb\dot{I}(x)}{1+bx}+ \ddot{I}(x)+ \dot{I}(x)^{2}\right]+ 2xC\left[\frac{as}{1+ax}+ \frac{b}{1+bx} + \dot{I}(x) \right]\nonumber\\
&&+\frac{C(a-b)ax-2\varepsilon x}{(1+ax)^{2}}.
\end{eqnarray}
\end{subequations}

\begin{acknowledgements}
PMT thanks the National Research Foundation and the University of
KwaZulu-Natal for financial support. SDM acknowledges that this
work is based upon research supported by the South African Research
Chair Initiative of the Department of Science and
Technology and the National Research Foundation. We are grateful to the referees for constructive comments
that have greatly improved the manuscript.
\end{acknowledgements}

\end{document}